\documentclass[12pt]{article}
\usepackage{epsf}
\usepackage{amsfonts}
\title{Notes  on the compatibility of type Ia supernovae data and varying--$G$ cosmology}
\author{F. Shojai and A. Shojai\\
Department of Physics, University of Tehran, Tehran, Iran.} 
\date{}
\usepackage{graphicx}
\begin{document}
\maketitle
\begin{abstract}
Observational data for type Ia supernovae, shows that the expansion of the universe is accelerated.  This accelerated expansion can be described by a cosmological constant or by dark energy models like quintessence. An interesting question may be raised here. Is it possible to describe the accelerated expansion of universe using varying--$G$ cosmological models? Here we shall show that the price for having accelerated expansion in slow--varying--$G$ models  (in which the dynamical terms of $G$ are ignored) is to have highly non--conserved matter and also that it is in contradiction with other data.
\end{abstract}
\section{Introduction}
Supernovae of type Ia are one of the best cosmological distance indicators. The measurement of their distance as a function of the redshift shows that the expansion of the universe is accelerating. For small redshifts  the apparent luminosity of type Ia supernovae is less than what would be expected in the hypothetical curvature dominated empty universe, while for high redshifts ($z>1.25$) the story is  converse. (See the results of the observations of Supernovae Cosmology Project\cite{SCP}, and High-z Supernovae Search Team\cite{HSST}) This means that the expansion of the universe is dominated by the dark energy component at late times.

One possible way for describing the supernovae data is to consider that coupling constants are not really constant on the cosmological scales and that can be regarded as fields. For the case of gravity theory we can have varying--$G$, varying--$c$, or varying--$\Lambda$  models\cite{mag}. In these models  either $G$ or $c$ is described by a scalar field and the equations of motion determine the dynamics of both the space--time metric and the \textit{varying} constant. In the context of the cosmological models, according to the cosmological principle these varying constants are only time--dependent.

In this paper we shall show that although one can describe the type Ia Supernovae by a slow--varying--$G$ model (a model in which the variation of $G$ is so small that its dynamical terms in the modified Einstein\rq{}s equations can be ignored), but the price is to have highly non--conserved matter. Therefore it is not possible to describe the data using regular conserved matter and a slow varying gravitational coupling constant. 
\section{Cosmological observations}
An important application of Einstein\rq{}s theory of General Relativity is the description of the universe in large scale. Observation of the redshifts of distant galaxies shows that the universe is expanding, and observations of cosmic microwave background radiation shows that the universe in large scale is homogenous and isotropic.

The geometry of a spatial homogeneous and isotropic space can simply be described by the FLRW metric given by:
\begin{equation}
ds^2=c^2 dt^2-a^2(t)\left ( \frac{dr^2}{1-kr^2}+r^2d\Omega \right)
\end{equation}  
We are not able to measure the distance of distant galaxies to us directly. The only thing we have at hand is the light emitted from the stars in that distant galaxies. We can simply measure the redshift and the flux of energy of the emitted light. The redshift is defined as 
\begin{equation}
z=\frac{\lambda_r-\lambda_e}{\lambda_e}
\end{equation}
and according to the above metric, it is related to the scale factor as: 
\begin{equation}
1+z=\frac{a(t_r)}{a(t_e)}
\end{equation}
where $t_r$ is the time that light is received to us and $t_e$ is the emission time.
Usually one normalize the scale factor $a(t_r)=a(t_0)=1$.

The energy flux is defined as
\begin{equation}
{\cal F}=\frac{{\cal L}}{4\pi (1+z)^2r^2}
\end{equation}
where $\cal L$ is the object\rq{}s luminosity and $r$ is the coordinate distance of the observed object with the observer. There are  two factors of $1/(1+z)$ in this relation. One because the frequency of the emitted light has redshift as a result of the expansion of universe. Another one  represents  time dilation effect in the  rate of radiation of energy. The luminosity distance is defined as:
\begin{equation}
d_{\cal L}=(1+z)r
\end{equation}
In astrophysics it is convenient to measure fluxes on a logarithmic scale called magnitude, $m$ defined by:
\begin{equation}
\frac{{\cal F}}{{\cal F}_0}=10^{-m/2.5}
\end{equation}
in which ${\cal F}_0$ is the corresponding flux of magnitude zero.
 Observations can determine the distance modulus defined as the difference between the object\rq{}s magnitude and its absolute magnitude, $M$ (the magnitude if the object were at 10 pc distance):
 \begin{equation}
\mu=m-M=5\log \left [\frac{(1+z)r}{10}\right ]
\label{7}
\end{equation}
where $r$ is measured in pc scale.

\begin{figure}[htp]
\centering
\includegraphics[scale=0.5]{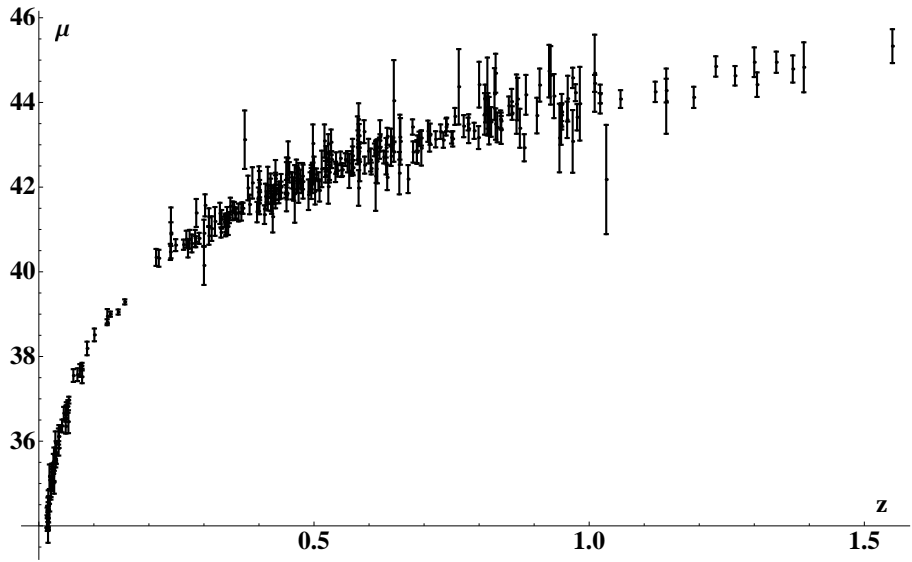}
\caption{Observational data for distance modulus as a function of redshift, for type Ia supernovae, using the compiled data of \cite{data}.}
\label{f0}
\end{figure}

  The observational data is best fitted on the theoretical predictions for a spatial flat space. Therefore the trajectory of light from the distant galaxy to us is governed by:
\begin{equation}
0=ds^2=c^2dt^2-a^2dr^2
\end{equation}
and thus:
\begin{equation}
r=\int_0^r dr\rq{}=c\int_t^{t_0}\frac{dt\rq{}}{a(t\rq{})}=c\int_{(1+z)^{-1}}^1\frac{da}{a\dot a}
\label{9}
\end{equation}
This enables us to express the distance modulus in terms of the redshift.
The observational data is shown in figure (\ref{f0}), using the compiled data in \cite{data}. 
\section{Varying--$G$ cosmology}
Since here we want to compare the $\Lambda$CDM model for which we have a constant $G$ and a cosmological constant $\Lambda$ with a model without cosmological constant but with a varying-$G$, it is essential to have the equations of motion for the general case of varying-$G$ model with cosmological constant.
   
The action functional for the Einstein\rq{}s theory of gravity is:
\begin{equation}
{\cal A}_{E-H}=\frac{1}{2\kappa_0}\int d^4x\sqrt{-g}(R+2\Lambda)+{\cal A}_m(\phi_m;g_{\mu\nu})
\end{equation}
in which $\kappa_0=8\pi G_0$ ($G_0$ is the present value of the gravitational constant and we set $c=1$) , $\Lambda$ is the cosmological constant, and ${\cal A}_m$ is the action for the matter fields $\phi_m$.

In order to have a varying--$G$ model, a simple way is to introduce a dimensionless field $\chi=G_0/G=\kappa_0/\kappa$ (where $G$ denotes the variable gravitational constant)  and include some dynamical terms for this new field. The action is thus:

\begin{equation}
{\cal A}=\frac{1}{2\kappa_0}\int d^4x \sqrt{-g}\left \{\chi(  R+2\Lambda)-Z(\chi)g^{\mu\nu}\partial_\mu\chi\partial_\nu\chi-2U(\chi)\right\}
+{\cal A}_m(\phi_m;g_{\mu\nu})
\label{ac}
\end{equation}  
The term in the bracket is the gravitational part of Lagrangian, in which the arbitrary function $Z(\chi)$  represent the coupling of the space--time metric with the scalar field and $U(\chi)$ is the potential term of the scalar field.

The equations of motion are then:
\begin{equation}
 G_{\mu\nu}+\Lambda g_{\mu\nu}=\kappa T_{\mu\nu}+\left \{ \frac{Z}{\chi}\left(  \nabla_\mu\chi\nabla_\nu\chi-\frac{1}{2}g_{\mu\nu}|\nabla\chi|^2   \right)+\frac{\nabla_\mu\nabla_\nu\chi}{\chi}-\frac{g_{\mu\nu}}{\chi}(\Box \chi+U)\right \}
 \label{MEE}
\end{equation}
\begin{equation}
2Z\Box\chi=-R-\frac{dZ}{d\chi}|\nabla\chi|^2+2\frac{dU}{d\chi}
\label{MEE2}
\end{equation}
The fact that the matter action is general covariant leads to 
\begin{equation}
\nabla_\mu T^{\mu\nu}=0
\label{EC}
\end{equation}
where the energy--momentum tensor is given by $\sqrt{-g}T^{\mu\nu}=2\delta {\cal A}_m/\delta g_{\mu\nu}$.

Before proceeding, it is useful to note that there are two pictures of such models. 
The above action  for the varying--$G$ models represents a picture  called \textit{Jordan picture} (usually the misleading term, Jordan frame is used). There is also another way of representing a varying--$G$ model via the following change of variables
\begin{equation}
\tilde g_{\mu\nu}=\chi g_{\mu\nu};\ \ \ \ \ 
\left (\frac{d\psi}{d\chi}\right )^2=\frac{3}{4\chi^2}+\frac{Z}{2\chi};\ \ \ \ \  
2V(\psi)=\frac{U(\chi)}{\chi^2};\ \ \ \ \ \tilde\Lambda=\frac{\Lambda}{\chi}
 \end{equation}
  The action in terms of these new variables is
  \begin{equation}
  {\cal A}=\frac{1}{2\kappa_0}\int d^4x \sqrt{-\tilde g}\left \{ \tilde R+2\tilde\Lambda-2\tilde g^{\mu\nu}\partial_\mu\psi\partial_\nu\psi-4V(\psi)\right\}
+{\cal A}_m(\phi_m;\tilde g_{\mu\nu}/\chi)
\label{eac}
  \end{equation}
This is called the \textit{Einstein picture}. The main difference is that in the Einstein picture the matter fields are non--minimally coupled to the metric.
This means that a varying--$G$ model  in Jordan picture produces a constant--$G$ model with non--minimal couplings  in Einstein picture. As a result in Einstein picture the gravitational constant is constant but particle masses are varying.
For a complete review on Einstein and Jordan pictures see \cite{far}.
In this paper we are using the Jordan frame.

It has to be noted that if the variation of the gravitational constant is small, the dynamical terms of $G$ may be ignored. These models can be called slow--varying--$G$ models (this may be done for any varying constant model, for varying--$c$ see \cite{a}). As a result slow varying--$G$ model is defined by Einstein equations with inserting a variable gravitational constant as an input on the right hand side. This may be justified by supposing that the extra terms (terms in the bracket) in the modified Einstein equations (\ref{MEE}) are ignorable with respect to the energy--momentum term.  This leads to the approximate equation:
\begin{equation}
\left ( G_{\mu\nu}+\Lambda g_{\mu\nu}\simeq\kappa T_{\mu\nu} \right )_{\textrm{\tiny slow varying--$G$}}
\end{equation}
Taking the divergence of this equation we have:
\begin{equation}
\left ( \nabla_\mu T^{\mu\nu}\simeq -T^{\mu\nu}\frac{\partial_\mu G}{G} \right )_{\textrm{\tiny slow varying--$G$}}
\end{equation}
in contradiction with the conservation law of energy--momentum (\ref{EC}). This is not surprising because the assumption of smallness of extra terms in modified Einstein equations is frame dependent. They can be small in one coordinate frame and large in another one. Therefore the general covariance would be broken and thus the energy--momentum is not conserved.

For our purpose it is necessary to apply this model to the flat FLRW solution. The equations would be:
\begin{equation}
3H^2=\kappa\rho+\Lambda+\left\{\frac{Z}{2\chi}\dot\chi^2-3H\frac{\dot\chi}{\chi}+\frac{U}{\chi}\right\}
\label{s1}
\end{equation}
\begin{equation}
2\dot H+3H^2=-\kappa p-\Lambda+\left\{ -\frac{Z}{2\chi}\dot\chi^2-\frac{\ddot\chi}{\chi}-2H\frac{\dot\chi}{\chi}+ \frac{U}{\chi}\right\}
\label{s2}
\end{equation}
\begin{equation}
Z(\ddot\chi+3H\dot\chi)=3(\dot H+2H^2)-\frac{Z\rq{}}{2}\dot\chi^2-U\rq{}
\end{equation}
where $H=\dot a/a$ is the Hubble parameter.
Combination of the first two equations leads to the conservation law $\dot\rho+3H(\rho+p)=0$, as it is expected and noted previously (see equation (\ref{EC})).

On the other hand, if one wants to use the above mentioned slow--varying--$G$ approximation and thus  choosing the gravitational coupling as an input not as a dynamical variable, then the equations would be:
\begin{equation}
\left (3H^2\simeq\kappa\rho+\Lambda  \right )_{\textrm{\tiny slow varying--$G$}}
\label{q1}
\end{equation}
\begin{equation}
\left ( 2\dot H+3H^2\simeq-\kappa p-\Lambda  \right )_{\textrm{\tiny slow varying--$G$}}
\label{q2}
\end{equation}
\begin{equation}
\left ( \dot\rho+3H(\rho+p)\simeq-\rho\frac{\dot\kappa}{\kappa}  \right )_{\textrm{\tiny slow varying--$G$}}
\label{q3}
\end{equation}
The condition for being able to go to this approximation is that the extra terms on the right hand side of the field equations (terms in bracket) (\ref{s1}) and (\ref{s2}), be negligible with respect to $\kappa\rho$ and $\kappa p$.  As a result for dust in which $p=0$ and when the cosmological constant is zero, this approximation is not physically acceptable.  We  shall come back to this point later in the conclusion.
\section{$\bf\Lambda$CDM v.s. Slow--Varying--$G$}
It is a well known fact that the cosmological observations can be  described by $\Lambda$CDM model.  In this model  the universe content is  dust and cosmological constant. 
A question may be raised here. Is it possible to describe  the observations by a slow-varying--$G$ model without any cosmological constant? To answer to this question consider the slow--varying--$G$ model (equations (\ref{q1})--(\ref{q3})) and define
\begin{equation}
\kappa=\kappa_0f(a);\ \ \ \ \ 
\Omega_M(a)=\frac{\kappa_0\rho(a)}{3H_0^2};\ \ \ \ \   
\Omega_\Lambda=\frac{\Lambda}{3H_0^2}=\textit{const.};\ \ \ \ \  
 \Omega=f(a)\Omega_M+\Omega_\Lambda
\label{25}
\end{equation}
then, we have
\begin{equation}
H^2=H_0^2(f(a)\Omega_M(a)+\Omega_\Lambda)
\label{26}
\end{equation}
and considering matter as dust, $p=0$, the equations (\ref{q1})--(\ref{q3}) would be
\begin{equation}
H^2=H_0^2\Omega
\label{27}
\end{equation}
\begin{equation}
\dot\Omega_M+3H\Omega_M=-\Omega_M\frac{f\rq{}}{f}aH
\end{equation}
where a prime over any quantity represents differentiating with respect to the scale factor. As a result
\begin{equation}
\dot\Omega+3H\Omega=-3H\Omega_\Lambda
\label{29}
\end{equation}
Two cases can be distinguished. First, if the gravitational constant is constant (i.e. $f=1$). The above equations can be solved to get
\begin{equation}
\Omega^{(1)}(a)=\Omega_M^{(1)}(1)(a^{-3}-1)+1
\end{equation}
and according to the observational data\cite{data} we have to set $\Omega_M^{(1)}(1)\simeq 0.287$.

\begin{figure}[htp]
\centering
\includegraphics[scale=0.5]{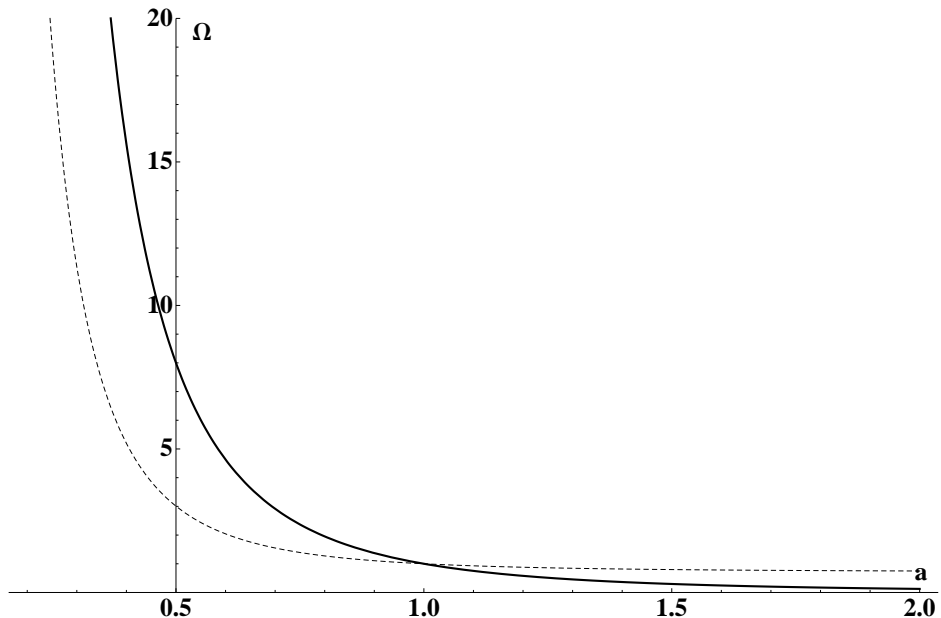}
\caption{$\Omega^{(1)}$ (dotted line) and $\Omega^{(2)}$ (thick line) as a function of scale factor.}
\label{f1}
\end{figure}

The second case is a slow--varying--G model without any cosmological constant. The solution is
\begin{equation}
\Omega^{(2)}(a)=a^{-3}
\end{equation}
First of all note that these two cases do not lead to the same dynamic as it is plotted in figure (\ref{f1}). Only in the vicinity of $a=1$ the two models are approximately the same.
However these two models are not so different in giving the dependence of distance modulus in terms of the redshift.
Using forms of $\Omega^{(1)}$ and $\Omega^{(2)}$ and integrating the relation (\ref{9}) and using equation (\ref{7}), one can obtain this dependence. 
The result is:
\begin{equation}
\mu^{(1)}=\mu_0+5\log\left [  -1.133+1.184 (1+z)\ \  _2F_1\left (\frac{1}{3},\frac{1}{2},\frac{4}{3},-0.403 (1+z)^3\right )\right ]
\end{equation}
\begin{equation}
\mu^{(2)}=\mu_0+5\log\left [ 2(1+z-\sqrt{1+z}) \right ]
\end{equation}
where $\mu_0$  have to be determined for best fitting to observational data. This is because of the fact that the absolute magnitude of a supernova cannot be determined exactly. As a result data only determines the shape of dependence of distance modulus on the redshift.

\begin{figure}[htp]
\centering
\includegraphics[scale=0.49]{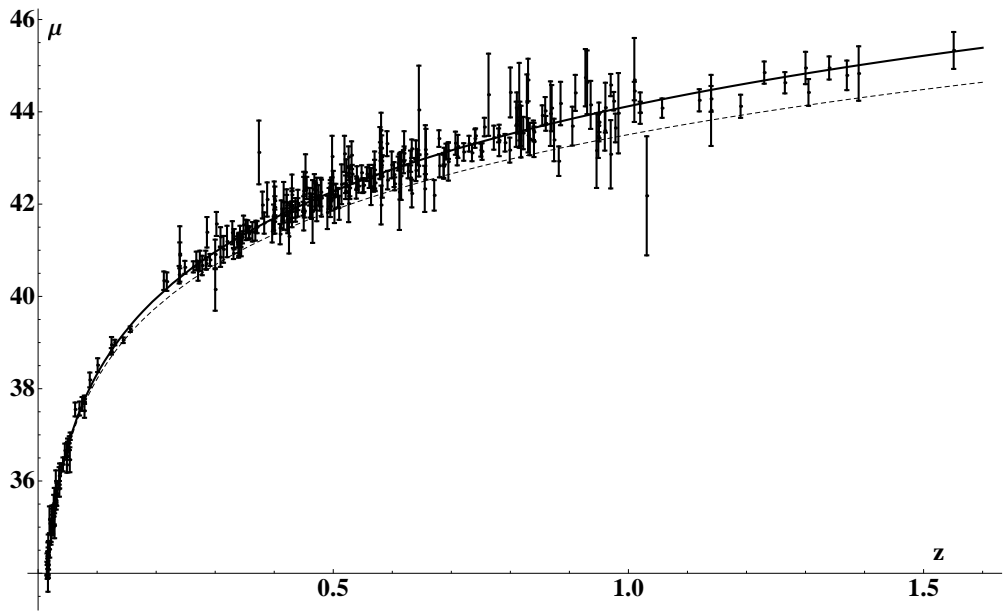}
\caption{Distance modulus in terms of the redshift for the two models ($\Lambda$CDM model (thick line) and slow--varying--$G$ model (dotted line)), as compared with the observational data. }
\label{f2}
\end{figure}

The above functions and observations are compared
in figure (\ref{f2}). Note that although the slow--varying-$G$ is somehow under fitted to the observational data, but noting the fact that we are talking about a very simplified model, one can conclude that the model can not be ruled out. The difference between the two models is larger for higher redshifts. This corresponds to the larger difference between the two $\Omega$\rq{}s at small scale factors (see figure (\ref{f1})).
The two models are more or less faithful to observations.

The price paid for the second model is not only the variation of $G$ but also the non--conservation of matter. If we assume that the matter is approximately conserved, we have to assume that the right hand side term in the conservation equation of matter, equation (\ref{q3}), can be ignored. Then the condition of having approximately the same dynamics is
\begin{equation}
\Omega_M^{(1)}(1)a^{-3}+(1-\Omega_M^{(1)})\simeq f(a)\Omega_M^{(2)}(a)
\end{equation}
and considering the conservation of matter in the second case, we get
\begin{equation}
\Omega_M^{(2)}(a)=\Omega_M^{(2)}(1)a^{-3}
\end{equation}
leading to
\begin{equation}
f(a)\simeq \frac{\Omega_M^{(1)}(1)}{\Omega_M^{(2)}(1)}\left( 1+ \frac{1-\Omega_M^{(1)}(1)}{\Omega_M^{(1)}(1)}a^3 \right)
\label{fa}
\end{equation}

This description, i.e. having slow--varying--$G$ without any cosmological constant (and dark matter) and having conserved matter is not necessarily consistent. To check the consistency, we have to check whether the equation (\ref{29}) (with $\Omega_\Lambda=0$) is approximately correct or not. Thus
the condition of consistency of this description is that
\begin{equation}
\dot\Omega^{(2)}+3H\Omega^{(2)}\simeq 0
\end{equation}
Normalizing this equation with $H_0$ and using relations in (\ref{25}) and defining
\begin{equation}
\eta(z)=3[1-\Omega_M^{(1)}(1)]\sqrt{\Omega_M^{(1)}(1)(1+z)^3+(1-\Omega_M^{(1)}(1))}
\end{equation}
the condition is that
\begin{equation}
\eta\sim 0
 \end{equation} 
 
\begin{figure}[htp]
\centering
\includegraphics[scale=0.5]{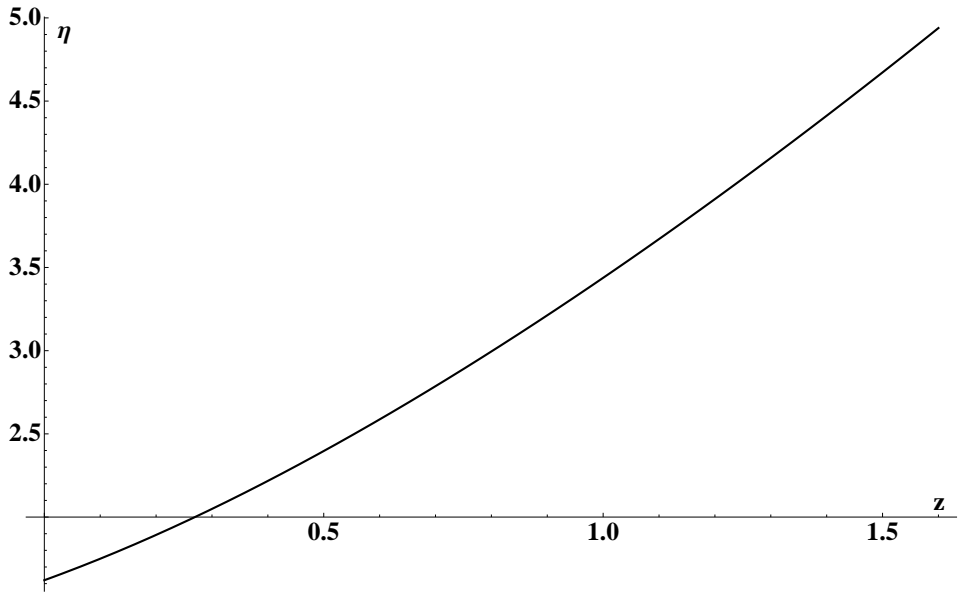}
\caption{The consistency function as a function of the redshift.}
\label{f3}
\end{figure}

 The function $\eta(z)$ is plotted in figure (\ref{f3}).
As it can be seen, this condition is highly violated. Therefore to have a slow--varying--$G$ model consistent with the observational data, matter should be highly non--conserved. Ignoring this fact can lead to incorrect conclusion that slow--varying--$G$ model ($G$ as input not as field) with conserved matter can describe the supernovae data as it is concluded in \cite{khar}.  Slow--varying--$G$ model has theoretical conflict with conservation of matter.  

It is necessary to check the compatibility of our results with other data about variation of the gravitational constant. 
From equation (\ref{fa}), one can evaluate the time variation of $G$ as:
\begin{equation}
\frac{\dot G}{G}=\frac{\dot f}{f}=\frac{3a^3\frac{1-\Omega_M^{(1)}(1)}{\Omega_M^{(1)}(1)}H}{1+a^3\frac{1-\Omega_M^{(1)}(1)}{\Omega_M^{(1)}(1)}}
\end{equation}
leading to 
\begin{equation}
\left (\frac{\dot G}{G}\right )_{\textrm{now}}\simeq 7\times 10^{-11} \textrm{ per year}
\end{equation}
The analysis of a long list of the observational data\cite{x} leads to an upper bound in the range between $1.5\times 10^{-10} \textrm{ per year}$ and $3\times 10^{-13} \textrm{ per year}$. Our value of $\dot G/G$ is somewhere within the upper bounds given by various observations. 
In fact a look at detailed data and discussions in \cite{x} shows that the above value for relative change of the gravitational constant can be only in agreement with constraints from pulsar stars. Some constraints (Earth--moon system, $\cdots$) are marginally violated. All the other constraints including terrestrial, radar ranging, globular clusters, stellar evolution, etc. are violated. 
As a result one can safely concludes that the slow--varying--$G$ model contradicts with the observational constraints on the value of relative change of $G$ and thus it is ruled out by the observations.   

The above conclusion only means that the slow varying regime is not acceptable and the dynamical terms should be considered. The dynamical terms can be added as it is done in (\ref{ac}), in Jordan picture, leading to the equations of motion (\ref{MEE}) and (\ref{MEE2}). As it is noted before, one can also use the  Einstein picture action (\ref{eac}). In principle, the potential term $U$ can be chosen such that the variation of $G$ has no contradiction at least with some of the observational data.
\section{Conclusions}
In this paper we considered the following question.
 \textit{Is it possible to explain the type Ia supernovae data using slow--varying--$G$ models instead of $\it \Lambda$CDM model}.  It is shown  that although this is possible,  the price paid is to have highly non--conserved matter. Moreover the value obtained for the relative variation of $G$ is in contradiction with other data and thus the slow--varying--$G$ model is ruled out by these data.

Another important problem is that if one wants to use slow--varying--$G$ model instead of $\Lambda$CDM, one has to check that if the condition of consistency of slow--varying--$G$ is satisfied or not. To have the consistency, the dynamical term of $G$ in equations (\ref{s1}) and (\ref{s2}) should be ignored with respect to energy--momentum tensor terms. As a result, if in the equation (\ref{s2}) we set $\Lambda=0$,  for dust ($p=0$), it is impossible to ignore dynamical terms for $G$. This means that also theoretically it is not accepted to describe the accelerated expansion only using a slow--varying--$G$ model.

On the other hand if we consider dynamical terms of the gravitational coupling constant, we are dealing with a scalar tensor theory for which with appropriate choice of potential term in the action one can describe the accelerated expansion, because it is equivalent to a quintessence model\cite{quin}.

\end{document}